\documentclass[a4paper,12pt]{article}
\usepackage{hyperref}
\usepackage{amsmath, amssymb, slashed, epsf, color, graphicx}
\usepackage{epsfig}
\newcommand{\lapprox}{%
\mathrel{%
\setbox0=\hbox{$<$}
\raise0.6ex\copy0\kern-\wd0
\lower0.65ex\hbox{$\sim$}
}}
\newcommand{\gapprox}{%
\mathrel{%
\setbox0=\hbox{$>$}
\raise0.6ex\copy0\kern-\wd0
\lower0.65ex\hbox{$\sim$}
}}

\newcommand{\ba}{\begin{array}}
\newcommand{\ea}{\end{array}}
\newcommand{\bd}{\begin{displaymath}}
\newcommand{\ed}{\end{displaymath}}
\newcommand{\be}{\begin{equation}}
\newcommand{\ee}{\end{equation}}
\newcommand{\bea}{\begin{eqnarray}}
\newcommand{\eea}{\end{eqnarray}}

\def\q2 {q^2}

\def\bt{\begin{table}}
\def\et{\end{table}}

\catcode`@=11 
\def \gsim{\mathrel{\mathpalette\@versim>}}
\def \lsim{\mathrel{\mathpalette\@versim<}}
\def \@versim#1#2{\lower0.4ex\vbox{\baselineskip\z@skip\lineskip\z@skip
     \lineskiplimit\z@\ialign{$\m@th#1\hfil##\hfil$%
     \crcr#2\crcr\sim\crcr}}}
\catcode`@=12 

\begin{document}

\begin{flushright}
{\small 
RECAPP-HRI-2013-016}
\end{flushright}

\begin{center}

{\large\bf Reconciling small radion vacuum expectation values with massive gravitons in an Einstein-Gauss-Bonnet
 warped geometry scenario}\\[15mm]
Ushoshi Maitra \footnote {E-mail: ushoshi@hri.res.in},
Biswarup Mukhopadhyaya \footnote{E-mail: biswarup@hri.res.in}\\
{\em Regional Centre for Accelerator-based Particle Physics \\
     Harish-Chandra Research Institute\\
Chhatnag Road, Jhusi, Allahabad - 211 019, India\\} 
Soumitra SenGupta \footnote{E-mail: tpssg@iacs.res.in} \\
{\em Department of Theoretical Physics,\\
 Indian Association for Cultivation of Science,\\
 2A and 2B Raja S.C. Mullick Road, Kolkata-700032, India}\\
[20mm] 
\end{center}
\abstract{In the usual 5-dimensional Randall-Sundrum scenario with warped geometry of the extra compact dimension,
the Goldberger-Wise mechanism for stabilisation of the radius of compactification can lead to a scalar field called
the radion. The radion can have implications in TeV-scale physics, which can be especially
noticeable if its vacuum expectation value (vev) is not far above a TeV. However a large mass of the 
first graviton excitation, which seems to be suggested by recent search limit, tends to make the radion vev,
far too large in the minimal model. We show that this is not the case if a Gauss-Bonnet term, containing higher powers 
of the curvature, is present in the 5-dimensional action. As a result, a radion with vev in the range 1-1.5 TeV can be consistent 
with the first graviton excitation mass well above 3 TeV.}   
\newpage

The phenomenology of warped compact spacelike extra dimensions has attracted considerable
attention for more than a decade now, mostly in the context of the Randall-Sundrum (RS) model ~\cite{Randall:1999ee}. Endowed
with an $S_1/Z_2$ orbifold symmetry for the extra dimension, and values of all bulk mass 
parameters close to the Planck mass, this scenario yields TeV-scale
mass parameters on one of the two branes kept at the orbifold fixed points. The process of
compactification leads to a tower of gravitons on this brane, popularly called the `visible brane'.
Since the massive modes of this tower have interaction to matter fields suppressed by only
a TeV-scale parameter, their detectability at the Large Hadron Collider (LHC) is being closely
studied, from both the experimental and theoretical angles.~\cite{ATLAS:2013,Chatrchyan:2013qha,
Das:2013lqa,YaserAyazi:2011at,Antipin:2007pi,Shivaji:2011re,Atag:2010bh,Xiao-Zhou:2013uqa,Murayama:2009jz, Randall:2008xg,Abazov:2005pi,
Allanach:2002gn,Tang:2012pv}

In the minimal RS model~\cite{Randall:1999ee}, the 5-dimensional background metric is
\begin{equation}
 ds^2 = e^{-2k_{RS} r_c \pi} \eta_{\mu\nu} dx^{\mu}dx^{\nu} + r_c^2 dy^2
\end{equation}
where $k_{RS} = \sqrt{\frac{-\Lambda}{24M^3}}$, $\Lambda$ is the Bulk cosmological constant, 
M is the 5-dimensional Planck mass and $r_c$ is the compactified radius. 
The 5-dimensional Planck mass is related to the reduced 4-dimensional Planck mass $(\bar{M_{Pl}})$ by the following manner 
\begin{eqnarray}
 \bar{M_{Pl}}^2  &=& \frac{M^3}{k_{RS}}(1 - e^{-2k_{RS} r_{c} \pi}) \\ \nonumber
  \bar{M_{Pl}}^2 &\simeq& \frac{M^3}{k_{RS}}
\end{eqnarray}
where $\bar{M_{Pl}} = 2 \times 10^{18}$ GeV.

The fraction $k_{RS}/\bar{M_{Pl}}$ is usually held to be well below unity, so that the classical solution is not
marred by yet unknown quantum effects in the trans-Planckian region. For $k_{RS}/\bar{M_{Pl}} = 0.1$, the observed limit  
on the mass of the first graviton KK excitation ($m_{G}$) has already gone up to about 2.4 - 2.6 TeV ~\cite{ATLAS:2013}. 
This sometimes raises the apprehension that  massive RS gravitons may elude us at the LHC.

However, another aspect of RS model is modulus stabilisation, which essentially consists in showing the radius $r_c$ of
the compact dimension as the vacuum expectation value (vev) of a modulus field. 
 The Goldberger-Wise mechanism is invoked in this context~\cite{Goldberger:1999uk,Goldberger:1999un}. The net consequence of this mechanism is the generation
of a 4-dimensional scalar potential involving a scalar field
called  the  radion. It can have substantial interaction with all standard model (SM) particles, 
which is stronger for smaller values of $\Lambda_{\varphi}$, the vev of the radion, which in 
turn is related to the vev of the modulus. The signature of the radion is a matter of interest at the LHC 
~\cite{Cho:2013mva,Ohno:2013sc,Chacko:2012vm,Barger:2011qn,Low:2012rj,Davoudiasl:2012xd,Barger:2011hu,Goncalves:2010dw,Cheung:2011nv}.
In addition, one can, in general, have mixing between the SM Higgs boson and the radion 
~\cite{Giudice:2000av,Chaichian:2001rq,Dominici:2002jv,Csaki:2000zn,Csaki:1999mp}.
The implication of this in the light of the LHC data has been the subject of recent studies
~\cite{Grzadkowski:2012ng,Kubota:2012in,deSandes:2011zs,Desai:2013pga}. 
On the whole the radion affects low-energy phenomenology more for smaller values of $\Lambda_{\varphi}$.
However, $\Lambda_{\varphi}$ increases concomitantly with $m_{G}$ in the minimal RS model, 
which again suggests less noticeable participation of the radion in collider phenomenology.

Here we explore a possibility of reconciling large $m_{G}$ (2.5 -- 10 TeV) with small 
 $\Lambda_{\varphi}$ (1.0 -- 1.5 TeV), if the 5-dimensional Einstein-Hilbert action is extended 
with a term, involving higher powers of the curvature. This brings back 
the hope of a rich radion phenomenology (with potential radion-Higgs mixing), 
even if the massive RS gravitons turn out to be too heavy to be seen with sufficient statistics.

While the RS model considers a 5-dimensional Einstein gravity action with a cosmological term,   
space-time dimension higher than four in general admits of  suitable combinations of 
higher order curvature terms when added to Einstein gravity still lead to second order field equations ~\cite{Brihaye:2011hf}.
In 5-dimensions, only a particular combination of the curvature terms, called the Gauss-Bonnet (GB) term gives rise to  the most general action for gravity.
Such a term appears as the correction at leading order in the inverse string tension, to Einstein gravity action in type II B string theory~\cite{Ohta:2010ae}.
Though a term of this type in the action turns out to be a trivial surface
term in 4-dimensions, it plays a crucial role in extra dimensional models. Such an addition can in principle modify phenomenological and cosmological signatures significantly and is 
therefore a subject of interest in recent times. The characteristic parameter of Einstein-Gauss-Bonnet (EGB) theory  is the coefficient
of the higher derivative terms, denoted here as $\alpha$. 
It has been shown that in the 
RS model, if a negative cosmological constant is added to
the 5-dimensional EGB  gravity action then the GB coupling 
parameter $\alpha$ can take values only within a limited interval.
In the context of warped phenomenology, the GB correction modifies the conventional RS model 
 by giving rise to an $\alpha$-dependent warp factor~\cite{Kim:1999dq, Kim:2000pz, Rizzo:2004rq, Choudhury:2013yg}. This in turn alters the correlation between $m_G$ and $\Lambda_{\varphi}$
such that a large  $m_G$ with small $\Lambda_{\varphi}$ may be obtained for an appropriate choice of $\alpha$ within the allowed range. 

The overall framework is described by the following action~\cite{Kim:1999dq, Kim:2000pz, Rizzo:2004rq, Choudhury:2013yg}:
\begin{eqnarray}
 S_{5} & = & S_{EH} + S_{GB} + S_{Brane} + S_{Bulk} \nonumber\\
 S_{EH} & = & \frac{M^{3}}{2}\int d^5 x \sqrt{-g_{(5)}} R_{(5)} \nonumber\\
 S_{GB} & = & \frac{\alpha M}{2}\int d^5 x \sqrt{-g_{(5)}} [R^{ABCD}_{(5)} R_{ABCD}^{(5)} - 4R^{AB}_{(5)} R_{AB}^{(5)} + R_{(5)}^2 ]\nonumber\\
 S_{Brane} & = & \int d^5 x \sum_{i=1}^{2}\sqrt{-g_{(5)}^{(i)}}[{\cal{L}}^{field}_{i} - T_{i}]\delta(y-y_{i})\nonumber\\
 S_{Bulk} & = & \int d^5 x \sqrt{-g_{(5)}}[{\cal{L}}^{field}_{Bulk} - 2\Lambda]
\end{eqnarray}
In the above action, $i$ is the Brane index, $i$=1(Hidden brane), 2(Visible brane) and ${\cal{L}}^{field}_{i}$ is the Lagrangian 
for the fields on the $i$th brane with the brane tension $T_{i}$. Similarly,  ${\cal{L}}^{Bulk}$ is the Lagrangian for the fields present in the bulk.

It has been shown in ~\cite{deBoer:2009pn} that the requirement of causality in the 
holographic dual boundary theory of EGB model puts an upper
bound on the magnitude of GB parameter that inturn yields
an appropriate viscosity-entropy ratio. Similar constraints on $\alpha$
have been derived in ~\cite{Choudhury:2013dia} in EGB theory
in presence of a Kalb-Ramond field.

As has been demonstrated in ~\cite{Randall:1999ee}, the 5-dimensional metric assumes the form
 \begin{equation}
  ds^2 = e^{-2A(y)}\eta_{\alpha \beta}dx^{\alpha}dx^{\beta} + r_c^2 dy^2
 \end{equation}

 On solving the 5-dimensional Einstein equation with the above ansatz, one obtains, to the 
leading order in $\alpha$, $A(y) = k_{\alpha}y$ where $k_{\alpha}$ is given by
\begin{equation}
 k_{\alpha} = \sqrt{\frac{3M^2}{16\alpha}[1-\sqrt{[1+4\frac{\alpha\Lambda}{9M^5}}]]}
\end{equation}
Integrating out the coordinate y from the 5-dimensional action, we arrive at the effective action
with the modified Planck scale 
\begin{equation}
  \bar{M_{Pl}}^2 \simeq \frac{M^3}{k_{\alpha}}
\end{equation}

In terms of $k_{RS}$ as defined below equation 1, we have 
\begin{equation}
 k_{\alpha} = \sqrt{\frac{3M^2}{16\alpha}[1-\sqrt{1-\frac{32 \alpha k_{RS}^2}{3M^2}]}}
\end{equation}
The reality of $k_{\alpha}$ demands an upper bound on $\alpha$, given by 
\begin{equation}
 \alpha \le \frac{3}{32}\frac{M^2}{k_{RS}^2}
\end{equation}
Combining equation[6] and [7], one can write 
\begin{equation}
 k_{\alpha}=\sqrt{\frac{3\bar{M_{Pl}}^{\frac{4}{3}}k_{\alpha}^{\frac{2}{3}}}{16\alpha}[1 - \sqrt{[1 - \frac{32 \alpha k_{RS}^2}{3 \bar{M_{Pl}}^{\frac{4}{3}} k_{\alpha}^{\frac{2}{3}}}]}]}
\end{equation}

\begin{figure}[htp]
\begin{center}
\includegraphics[scale=0.6]{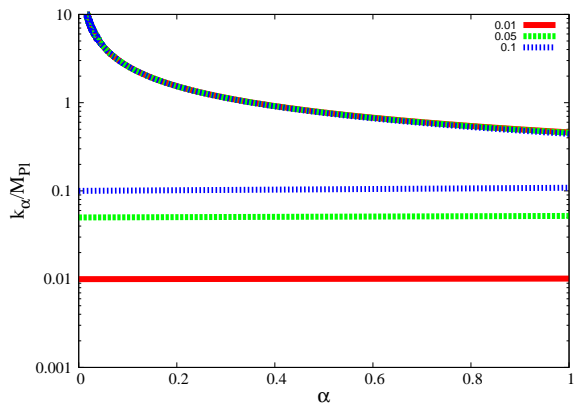} 
\caption{\footnotesize {Variation of $\frac{k_{\alpha}}{\bar{M_{Pl}}}$ with $\alpha$ 
for $\frac{k_{RS}}{\bar{M_{Pl}}}$ = 0.01(red), 0.05 (green)
and 0.1(blue).The upper branches are largely coincident for three different values of
$\frac{k_{RS}}{\bar{M_{Pl}}}$.}  \label{fig:x_alpha}}
\end{center}
\end{figure}

With this modified warp factor the Goldberger-Wise stabilization mechanism~\cite{Goldberger:1999uk,Goldberger:1999un} can be easily generalised to 
stabilize the modulus $r_c$ to its desired value for resolving the heirarchy problem.
From equation[9] the GB parameter $\alpha$ can be expressed as
\begin{equation}
 \alpha = \frac{3}{8}\frac{[(\frac{k_{\alpha}}{\bar{M_{Pl}}})^2 - (\frac{k_{RS}}{\bar{M_{Pl}}})^2]}{(\frac{k_{\alpha}}{\bar{M_{Pl}}})^{\frac{10}{3}}}
\end{equation}
The plot for $\frac{k_{\alpha}}{\bar{M_{Pl}}}$  against $\alpha$, for various $\frac{k_{RS}}{\bar{M_{Pl}}}$, is shown in Figure ~\ref{fig:x_alpha}.
For each $k_{RS}/\bar{M_{Pl}}$, there are two distinct branches, one of which yields the limit appropriate for $\alpha \rightarrow 0$, while the other
yields large values of $k_{\alpha}$ compared to $k_{RS}$. The branches that corresponds to $k_{\alpha} \simeq k_{RS}$ (the lower ones)
are rather distinct for the chosen values of $k_{RS}/\bar{M_{Pl}}$. On the other hand, they tend to merge when $k_{\alpha} >> k_{RS}$
(the upper branch), thus exhibiting very small dependence on $k_{RS}$. 

In the minimal RS model, the graviton KK mode masses are given by ~\cite{Davoudiasl:1999jd,Davoudiasl:2000wi,Chang:1999yn}
\begin{equation}
m_{n} = k_{RS}x_{n}e^{-k_{RS}r_c\pi}
\end{equation}
With the GB correction, the expression gets modified to ~\cite{Choudhury:2013yg}
\begin{equation}
m_n = k_{\alpha}x_ne^{-k_{\alpha}r_c\pi}
\end{equation}
where $m_{n}$ is the mass of the $n^{th}$ mode and $x_{n}$ is the nth root of $J_{1}(x)$.

It may be noted that despite the change in the value of the parameter 'k' due to 
the GB coupling $\alpha$, magnitude of the exponent is kept in the range 11.4-11.7,
in order to achieve a heirarchy between the Planck and electroweak scales. 
This in turn keeps the graviton KK mode 
coupling with the brane fields similar to that in RS model, with $k_{\alpha}$ replacing $k_{RS}$.
If we consider the first KK mode of RS graviton then its mass is given by
\begin{equation}
 m_G = k_{\alpha}e^{-k_{\alpha} r_c\pi}x_1
\end{equation}
where $x_1$ = 3.83. From the above relation, increase in $m_G$ implies increase in $k_{\alpha}$ for a fixed value of the warp factor.

Let us now recall the emergence of radion field.
The compactification radius, is set to a stable value $r_c$ by hand in the original RS model. 
However it is more satisfying to generate $r_c$ dynamically as the vev of a scalar field called
the modulus field, T(x). Thus, we have $ \langle T(x)\rangle \equiv r_c$. The modulus field
 with proper canonical redefinition gives 
a field $\varphi(x)$ which is the radion field with vev $\Lambda_{\varphi}$. The field T(x) can be
mapped to a standard scalar field $\varphi(x)$ (called the radion) with canonical kinetic energy term, 
via the redefinition ~\cite{Goldberger:1999uk,Goldberger:1999un}
\begin{equation}
 \varphi(x) = \Lambda_{\varphi}e^{-k_{\alpha}\pi(T(x)-r_c)} 
\end{equation}
Thus $\Lambda_{\varphi}$ emerges as the radion vev, being the value of $\varphi(x)$ for $T(x)=r_c$.
The radion turns out to be massive, with mass of the order of a TeV or less, when the effective radion
potential has the appropriate boundary conditions. This is the essence of the Goldberger-Wise mechanism,
with effects of the Gauss-Bonnet term taken into account. The substitutions necessary for obtaining
a canonical Lagrangian for $\varphi(x)$ lead to ~\cite{Giudice:2000av}
 
\begin{equation}
 \Lambda_{\varphi} = \sqrt{24}M_{Pl}e^{-k_{\alpha}r_{c}\pi} 
\end{equation}
The radion couples to the SM particles via the trace of the energy-momentum tensor $(T^{\mu}_{\nu})$:
\begin{equation}
 {\cal{L}}_{int} = \frac{\delta \varphi(x)}{\Lambda_{\varphi}}T^{\mu}_{\mu}
\end{equation}
where $\delta \varphi(x) = \varphi(x) - \Lambda_{\varphi}$ .

Thus all processes involving the interaction of the radion with SM fields, which are
at the root of all experimental signatures of the radion, have enhanced rates for relatively low
values of $\Lambda_{\varphi}$.

Thus $m_G$ with $\Lambda_{\varphi}$ are related as
\begin{equation}
 m_G = \Lambda_{\varphi}\frac{k_{\alpha}}{\bar{M_{Pl}}}\frac{x_1}{\sqrt{24}}
\end{equation}
For $\alpha \rightarrow 0$ limit we retreive our original RS limit.

From the above, we can see that, if we restrict ourself in the minimal RS model, then
an increase in $m_G$ leads to larger $\Lambda_{\varphi}$ for fixed $k_{RS}/\bar{M_{Pl}}$.
As the recent exclusion limits on $m_G$ have already gone up to 2.5 TeV ~\cite{ATLAS:2013} for $k_{RS}/\bar{M_{Pl}} \simeq 0.1$.
This will imply a minimum value of $\Lambda_{\varphi}$ of about 31 TeV. With $k_{RS}/\bar{M_Pl} \approx 1$, minimum 
value of $\Lambda_{\varphi}$ is about 5 TeV for $m_G \approx 3.5$ TeV. If we want smaller $\Lambda_{\varphi} (\simeq 1TeV)$
with high $m_{G}$, then it becomes necessary in the minimal RS model to have $k_{RS}$ in the trans-Planckian region.
But that in turn would make the classical solution questionable.

\begin{figure}[htp]
\begin{minipage}[b]{0.5\linewidth}
 \centering								
\includegraphics[scale=0.6]{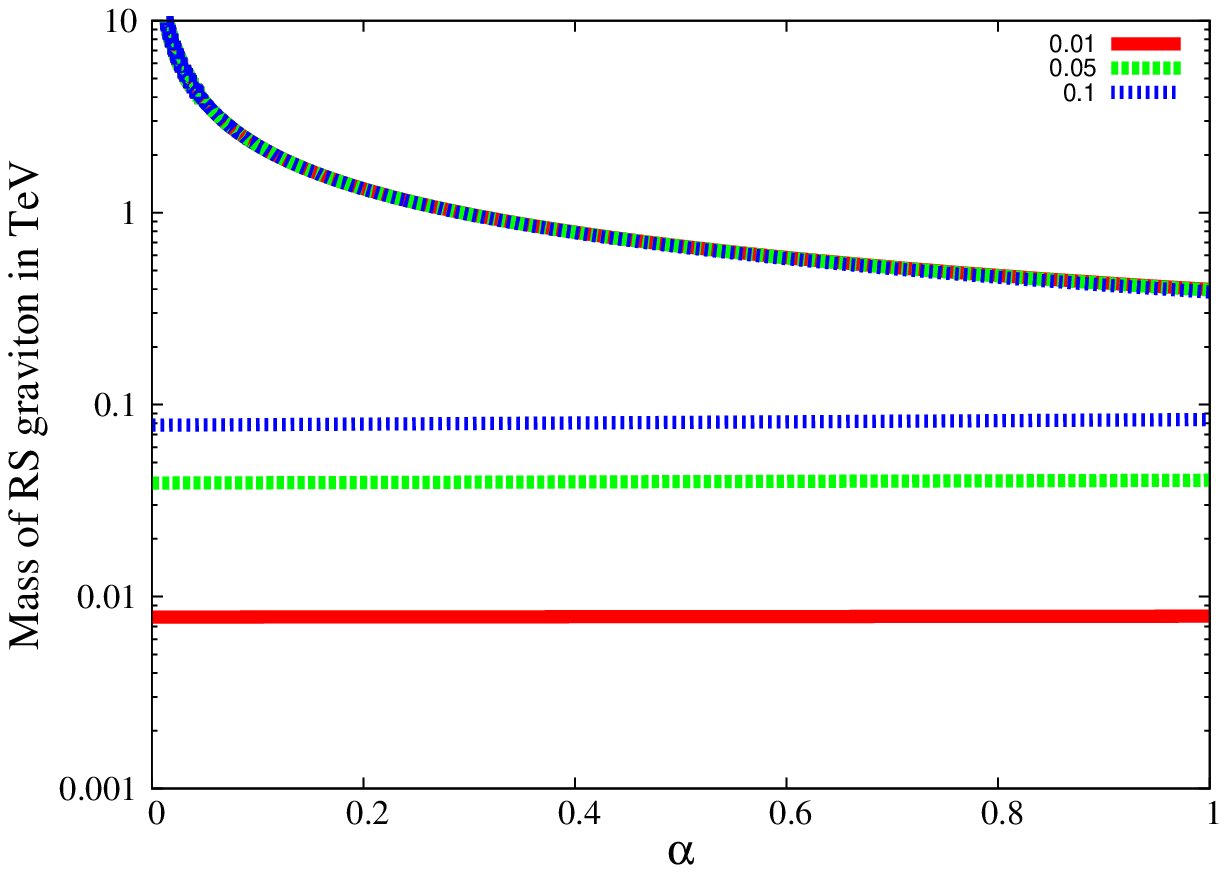} 					
\caption{\footnotesize {Variation of mass of the 1st KK mode 
of RS graviton with $\alpha$ for $\frac{k_{RS}}{\bar{M_{Pl}}}$ = 0.01(red), 0.05 (green)	
and 0.1(blue) with $\Lambda_{\varphi}$ = 1.0 TeV.
The upper branches are largely coincident for three different values of
$\frac{k_{RS}}{\bar{M_{Pl}}}$.}
 \label{fig:m_alpha_1}}
 \end{minipage}
\hspace{0.3cm}
\begin{minipage}[b]{0.5\linewidth}
 \centering
\includegraphics[scale=0.6]{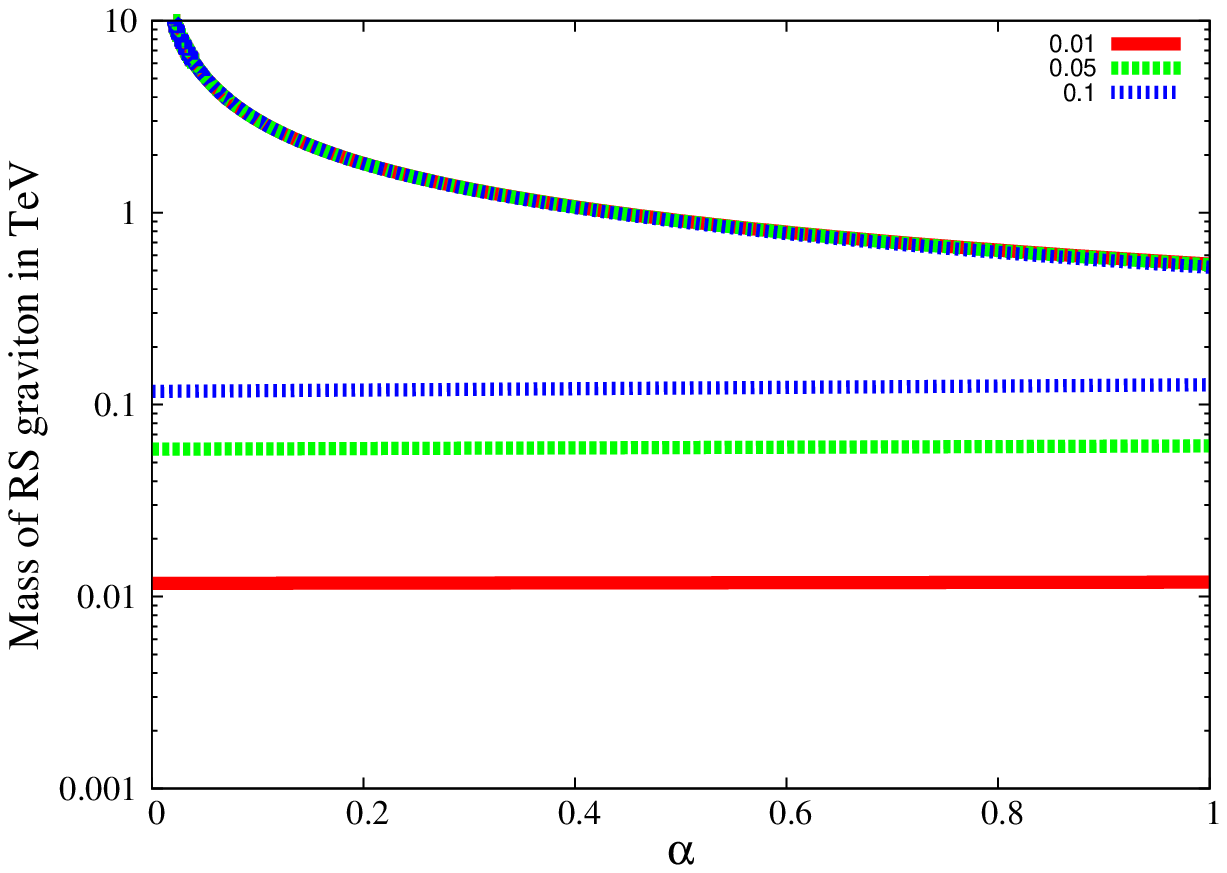} 
\caption{\footnotesize {Variation of mass of the 1st KK mode of RS graviton with $\alpha$ for $\frac{k_{RS}}{\bar{M_{Pl}}}$ = 0.01(red), 0.05 (green)
and 0.1(blue) with $\Lambda_{\varphi}$ = 1.5 TeV.
The upper branches are largely coincident for three different values of
$\frac{k_{RS}}{\bar{M_{Pl}}}$.} 
\label{fig:m_alpha_2}}
\end{minipage}
 \end{figure}

As has already been mentioned, a conciliatory mechanism is offered by the GB term. With its introduction, $k_{RS}$, which
is essentially related to the bulk cosmological constant, is replaced by the derived quantity $k_{\alpha}$
whose value is not restricted like $k_{RS}$. This may be observed from Figure[1] in the
upper branch, when, a large $k_{\alpha}/\bar{M_{Pl}}$ is obtained for smaller values of $k_{RS}/\bar{M_{Pl}}$.
In such a situation one has the liberty to have $\Lambda_{\varphi}$ in as low a range of 1-1.5 TeV. For a given
value of $\Lambda_{\varphi}, \alpha$ and $m_G$ are related by 
\begin{equation}
 \alpha = \frac{3}{8}(\frac{\Lambda_{\varphi}x_1}{\sqrt{24}})^{\frac{4}{3}}[\frac{m_{G}^2(\alpha) - m_{G}^2(\alpha = 0)}{m_G^{\frac{10}{3}}(\alpha)}]
\end{equation}
The plots for $m_G$ against $\alpha$ for various values of $\frac{k_{RS}}{\bar{M_{Pl}}}$ for
$\Lambda_{\varphi} = 1.0, 1.5 $ TeV are shown in Figure ~\ref{fig:m_alpha_1} and Figure ~\ref{fig:m_alpha_2}.
The figures indicate that for appropriate choices for the GB parameter $\alpha$ the mass of the 1st graviton KK mode can 
be well above the lower bound set by the recent ATLAS data ~\cite{ATLAS:2013} keeping the radion vev at a relatively small value of the order of $1-1.5$ TeV.
For example even for $k_{RS}/\bar{M_{PL}} = 0.1$ one can have $m_G$ as high as $3$ TeV for $\alpha = 0.1$. This value of $\alpha$
is consistent with all theoretical bound including that set by the holographic dual field theory of EGB
model.

To conclude, the introduction of a GB term can delink the phenomenology of the radion from that of the 
first excitation of graviton. As has been pointed out above, one may have values of the coefficient $\alpha$, 
which makes $m_G$ high enough to satisfy all the observed limits, with even the possibility of the graviton
signal evading detection at the LHC. Such values of $m_G$ and $\alpha$ can still be consistent with values of 
$\Lambda_{\varphi}$ in the range 1-1.5 TeV, so long as $r_c$ is appropriately set to yield a value of around 11.5
for the exponent in the warp factor. 

Extensive investigations in recent times \cite{Grzadkowski:2012ng,Kubota:2012in,deSandes:2011zs,Desai:2013pga} 
have revealed that radion-Higgs mixing 
can have bearing on the LHC Higgs data in various channels. The data on the WW*, ZZ*, $\gamma \gamma$ and $b\bar{b}$
channels for example, have been found to yield constraints on the parameter space spanned by the radion mass, its vev
$\Lambda_{\varphi}$, and the degree of radion-Higgs mixing. The effect of the radion-higgs mixing in such an analysis is more for smaller
$\Lambda_{\varphi}$  which, in the minimal RS scenario, would be highly disfavored by the graviton mass limit. While this 
has often prompted the postulation of $\Lambda_{\varphi}$ as high as 5-10 TeV, our result reopens the prospect of much
lower $\Lambda_{\varphi}$ for large graviton masses. Thus the existence of a GB term in the 5-dimensional action may lead
to a situation where a radion rather than a KK graviton is the harbinger of the RS framework, buttressed with the 
Goldberger-Wise mechanism.

{\bf{Acknowledgements:}} This work was partially supported by
funding available from the Department of Atomic Energy, Government of
India for the Regional centre for Accelerator-based Particle Physics,
Harish-Chandra Research Institute(RECAPP). SSG would like to thank
RECAPP for hospitality in the earlier phase of the study. UM and BM 
thank the Indian Association for the Cultivation of
Science, Kolkata for hospitality at the final stage of the project.

\end{document}